\newcommand{\nn}{\nonumber\\}

\newcommand{\be}{\begin{eqnarray}}
\newcommand{\ee}{\end{eqnarray}}

\newcommand{\bea}{\begin{eqnarray}}
\newcommand{\eea}{\end{eqnarray}}
\newcommand{\beas}{\begin{eqnarray*}}
\newcommand{\eeas}{\end{eqnarray*}}
\documentclass[twocolumn,aps,showpacs]{revtex4}
\usepackage{color}
\usepackage{graphics}
\usepackage{epsfig}
\usepackage{amsmath}
\usepackage{amssymb}
\usepackage{amsthm}
\begin{document}
\title{Chiral phase transition in relativistic heavy-ion
collisions with weak magnetic fields: ring diagrams in the linear sigma model}
\author{Alejandro Ayala$^{1,2,3}$, Adnan Bashir$^2$, Alfredo
  Raya$^2$ and Angel S\'anchez$^2$}   
\affiliation{$^1$Instituto de Ciencias Nucleares, Universidad
Nacional Aut\'onoma de M\'exico, Apartado Postal 70-543, M\'exico
Distrito Federal 04510, M\'exico.\\
$^2$Instituto de F\1sica y Matem\'aticas,
Universidad Michoacana de San Nicol\'as de Hidalgo, Apartado Postal
2-82, Morelia, Michoac\'an 58040, M\'exico.\\
$^3$Centro Brasileiro de Pesquisas F\1sicas,CBPF-DCP,
Rua Dr. Xavier Sigaud 150, 22290-180, Rio de Janeiro, Brazil.}

\begin{abstract} 

Working in the linear sigma model with quarks, we compute the
finite-temperature effective potential in the presence of a weak magnetic
field, including the contribution of the pion ring diagrams and considering
the sigma as a classical field. In the approximation where the pion
self-energy is computed perturbatively, we show that there is a region of the
parameter space where the effect of the ring diagrams is to preclude the phase
transition from happening. Inclusion of the magnetic field has small effects
that however become more important as the system evolves to the lowest
temperatures allowed in the analysis.

\end{abstract}

\pacs{11.10.Wx; 11.30.Rd; 12.38.Mh; 25.75.Nq}

\maketitle

\date{\today}

\section{Introduction}

In recent years it has been possible to produce and study hadronic matter at
high densities and temperatures by means of collisions of heavy nuclei at high
energies~\cite{PoSLHC07}. There are convincing signals that reveal the
production of deconfined matter where the degrees of freedom involved are the
quarks and gluons of QCD, forming the so called quark-gluon plasma (QGP). A
common feature of these signals is their strengthening as the centrality of
the collision increases. 

At the same time it has been realized that a host of new phenomena can also
happen for not so central collisions. Among these, it has been
pointed out that for peripheral collisions, a magnetic field of a
non-negligible strength is generated~\cite{Kharzeev}. The origin of this field
is two-fold: On one hand, in non-central collisions, there is a local
imbalance in the momentum carried by the colliding nucleons in the
target and projectile that generates a non-vanishing local angular 
momentum~\cite{Wang, Becattini} which in turn 
produces a magnetic field, given the net positive charge present in the
collision. On the other hand, the spectator nucleons can be thought of as
currents of net positive charge moving in opposite, off center, directions
which in turn produce a magnetic field that adds up in the interaction
region. 

An interesting question that emerges from this scenario is whether a magnetic
field can influence the phase transitions that may 
occur during the reaction, in particular the chiral phase transition. There
are known examples where magnetic fields are able to change the nature of a
phase transitions. Most notably is the Meissner effect where the phase
transition of type I superconductors changes from second to first order
when an external magnetic field is applied. Magnetic catalysis is another
phenomenon whereby a magnetic field is able to dynamically generate
masses in QED, regardless of the strength of the field~\cite{Gusynin}.
More recently, it has been shown that in the presence of primordial magnetic
fields, the electroweak phase transition, that took place in the early
universe for temperatures of order 100 GeV, gets also
strengthened~\cite{Ayala}. In a similar connection, the dynamical
generation of anomalous magnetic moment of the electron has also be
unveiled in Ref.~\cite{VivianEfrain}.

Calculations of the intensity of the field produced in this kind of collisions
show that for very early proper times after the reaction ($\tau\lesssim 0.1$
fm) the field reaches values $eB\simeq 6 (m_\pi^{vac})^2$, where $m_\pi^{vac}$
is the vacuum pion mass, even for mid-peripheral collisions. The intensity
decreases with the proper time $\tau$ as $eB\propto 1/\tau^3$ in such a way
that for $\tau\simeq 1$ fm, namely, for times when the standard picture of a
heavy-ion reaction places the existence of the equilibrated QGP, $eB\lesssim 
0.1\ (m_\pi^{vac})^2$~\cite{Kharzeev}, that is, already two orders of magnitude
smaller than at the very early stages of the collision.

In a recent work, the chiral phase transition in relativistic heavy-ion
collisions has been examined in the presence of strong magnetic fields using
the linear sigma model~\cite{Fraga}. Working with the hierarchy of scales where
$(m_\pi^{vac})^2\ll T^2 \ll eB$, with $T$ being the temperature around the phase 
transition, the authors conclude that the effect is to turn a
  crossover into a weak first order transition, a result that might be
  relevant even for the physics of the 
primordial QCD transition. Nevertheless, as mentioned
above, a more realistic scenario in heavy-ion collisions
should be to consider that for the times when 
the initial chromoelectric fields decohere in the aftermath of the collision
and give rise to partons --which in turn are the appropriate degrees of
freedom to describe the chiral phase transition-- the magnetic field in the
interaction region might not be that strong. At these times the hierarchy of
scales is such that the magnetic field is the smallest of all and the
deconfinement/chiral phase transition temperature is the largest one, while
the pion vacuum mass, occupies an intermediate place. Furthermore, the
analysis of Ref.~\cite{Fraga}  neglects the contribution from the so called
ring diagrams which are known to be important at high temperatures to account
for the infrared properties of the 
plasma~\cite{Carrington, Petropoulos}. Similar considerations, in QED for
instance, have already been reported~\cite{RingQED}. In the context of the QGP
and hadron matter, the 
effects of magnetic fields have also been recently looked at. Such studies
include their influence on making evident possible topological-charge
transitions in heavy-ion collisions~\cite{Kharzeev, Fukushima}, on the
confinement/deconfinement phase transition --in the Abelian approximation of
the chromomagnetic field--~\cite{Agasian} and on the hadron
structure~\cite{Tiburzi}.

In this work we undertake the calculation of the effective potential at finite
temperature, in the presence of weak magnetic fields. We use the linear sigma
model as the working tool to describe the chiral phase transition in
relativistic heavy-ion collisions. We work explicitly with the hierarchy of
energy scales where $eB\ll m^2\ll T^2$, with $m$ being a generic mass
appearing in the calculation and considering that the
interaction region is subject to an external magnetic field directed along the
positive $z$-axis. We work up to the contributions of the ring diagrams since
for the aforementioned hierarchy of scales, the effects of the magnetic
fields appear only at this level. Computation of the ring diagrams calls for
the calculation of the pion finite temperature self-energy in the presence of
the field. We approximate this self-energy by its perturbative value although,
for the large couplings involved, the calculation most likely overshoots the
true result. We parameterize the lack of an accurate non-perturbative treatment
introducing a parameter to control the strength of the self-energy and analyze
the consequences in this parameter space. To incorporate the weak magnetic
field we use the Schwinger proper time method to write the charged particle
propagators in a weak field expansion.  

The work is organized as follows: After a brief summary in Sec.~\ref{secII} of
the linear sigma model Lagrangian, we perform the calculation of the finite
temperature effective potential in Sec.~\ref{secIII}, considering that the
sigma field is classical and going up to the contribution of the ring
diagrams in the presence of a weak magnetic field. In Sec.~\ref{secIV} we
present the numerical results of the work and show that in a certain region of
the parameter space, even in the absence of
magnetic fields, the ring diagrams preclude the development of the phase
transition. Inclusion of the weak magnetic field has little impact on the phase
transition. Finally in Sec.~\ref{secV} we discuss our results and present the
conclusions.

\section{The linear sigma model with quarks}\label{secII}

The Lagrangian for the linear sigma model is given by
\be
   {\mathcal {L}}&=&\frac{1}{2}(\partial_\mu\sigma)^2 +
   \frac{1}{2}(\partial_\mu{\bf{\pi}})^2 + 
   \frac{\mu^2}{2}(\sigma^2+{\bf{\pi}}^2)\nn 
   &-&
   \frac{\lambda}{4}(\sigma^2+{\bf{\pi}}^2)^2 + 
   i\bar{\psi}\gamma^\mu\partial_\mu\psi\nn 
   &-&
   ig\bar{\psi}{\bf{\tau}}\gamma_5\psi��\cdot{\bf{\pi}} -
   g\bar{\psi}\psi\sigma,
\label{laglsm}
\ee
where $\psi$ is a SU(2) isospin doublet of massless quarks,
${\bf{\pi}}=(\pi_1,\pi_2,\pi_3)$ is an isospin triplet representing the pions
and $\sigma$ is an isospin singlet.

When the mass parameter $\mu^2$ is positive, the Lagrangian admits
a broken symmetry vacuum solution given by the minimum of the
classical potential
\be
   V^{(cl)}=-\frac{\mu^2}{2}(\sigma^2+{\bf{\pi}}^2)+
   \frac{\lambda}{4}(\sigma^2+{\bf{\pi}}^2)^2.
\label{Vcl}
\ee
Choosing this minimum along the $\sigma$ direction, the {\it vacuum
expectation values} for the sigma and pion fields are given by  
\be
   \langle\sigma\rangle&=&v_0
   \equiv \mu / \sqrt{\lambda}\nn
   \langle{\bf {\pi}}\rangle&=&0.
\label{vev}
\ee
$v_0$ is also called the classical vacuum which minimizes the action for
uniform field configurations. To study the quantum properties of the system,
we define the shifted field $\sigma'$ by 
\be
   \sigma = v + \sigma',
\label{shift}
\ee
where $v$ is taken as a variable. When $v=v_0$, $\sigma'$ represents the field
configuration around the classical vacuum. 

As a function of the shifted field, after symmetry breaking, the Lagrangian of
Eq.~(\ref{laglsm}) becomes a theory describing a $\sigma'$ field, three pion
fields and a quark-doublet field with masses given by 
\be
   m_{\sigma'}^2(v)&=&3\lambda v^2 - \mu^2\nn
   m_{\pi}^2(v)&=&\lambda v^2 - \mu^2\nn
   m_q(v)&=&gv.
\label{masses}
\ee
Armed with the fundamentals of the linear sigma model, we proceed to compute
the effective potential.

\section{The effective potential}\label{secIII}

\subsection{Tree and one-loop}\label{secIIIa}

The tree level potential is given by
\be
   V^{(tree)}=-\frac{\mu^2}{2}v^2 + \frac{\lambda}{4}v^4.
\label{Vtree}
\ee
We consider that the $\sigma'$ field is very heavy and thus treat it only
classically. To one-loop order, the contribution to the effective potential in
the imaginary-time formulation of thermal field theory is given, for bosons,
by 
\be
   V^{(1)}_{b}={\mathrm{s}}_bT\sum_n\int\frac{d^3k}{(2\pi)^3}\ln(D^{-1})^{1/2},
\label{Vbos1loop}
\ee
whereas for fermions, by
\be
   V^{(1)}_{f}={\mathrm{s}}_fT\sum_n\int\frac{d^3k}{(2\pi)^3}{\mbox{Tr}}
   \ln(S^{-1}),
\label{Vfer1loop}
\ee
where s$_{b,f}$ are the degeneracy factors accounting for the internal degrees
of freedom for bosons (isospin) and fermions (isospin and color),
respectively, $n$ is the index for the Matsubara frequency and $D$ and $S$
represent the boson and fermion Matsubara propagators. For
charged particles, these propagators should include the effect of the external
magnetic field. Nevertheless, it has been shown~\cite{Ayala} that for the
hierarchy of energy scales considered, the terms containing the effects of the
magnetic field are subdominant. 

Equations~(\ref{Vbos1loop}) and~(\ref{Vfer1loop}) contain both a vacuum and a
finite temperature piece. The vacuum piece exhibits the usual ultraviolet
divergence that needs to be removed. In the present context, this means
that, up to additive constants, the effective potential should contain only
finite $v$-dependent terms. This is accomplished, for instance, by introducing
counter-terms to absorb the infinities. The usual physical conditions
implemented to fix the counter-terms require that the position of the minimum
of the effective potential, as well as the mass of the $\sigma'$ field
maintain their classical values~\cite{Das}. However, for theories with
massless modes --such as the pions in the present case, whose mass
vanishes at $v=v_0$-- this procedure breaks down. The problem is that in order
to get the mass of the $\sigma'$ field from the effective potential one
requires computing the second derivative, since this mass is the inertia along
the $\sigma'$-axis. This derivative turns out to not be defined at such
value. Thus, the second condition needs to be replaced by another appropriate
one in a manner that we proceed to explain.

First, let us introduce the general expression for the one-loop renormalized
effective potential, where we add the counterterms to absorb the
$v$-dependent infinities 
\be
   V^{(1)}_{ren}&=&-\frac{\mu^2}{2}v^2 + \frac{\lambda}{4}v^4 
               + \left(\frac{a(\Lambda) - \delta\mu^2}{2}\right)v^2
   \nonumber\\
               &+& \left(\frac{b(\Lambda) + \delta\lambda}{4}\right)v^4
   + 3 I(m_\pi,\Lambda)
   \nonumber\\
               &-& 24I(m_f,\Lambda).
\label{Veffrenor}
\ee
The last two terms account for the pion and fermion contributions to the
vacuum effective potential, respectively, which from Eqs.~(\ref{Vbos1loop})
and~(\ref{Vfer1loop}) involve the ultraviolet cutoff $(\Lambda)$ dependent
function $I(m,\Lambda)$ defined as
\be
   I(m,\Lambda)=\frac{1}{2\pi^2}\int_0^\Lambda dk k^2\sqrt{k^2+m^2}.
\label{defI}
\ee
The counter-terms in Eq.~(\ref{Veffrenor}), $a(\Lambda)$ and $b(\Lambda)$, are
introduced to take care of the $v$-dependent infinities, whereas the
counter-terms $\delta\mu^2$ and $\delta\lambda$, account for finite terms that
might shift the coefficients of the $v^2$ and $v^4$ terms, respectively.

After some straightforward algebra where the $v$-dependent infinities are
absorbed, the one-loop renormalized effective potential becomes
\be
   V^{(1)}_{ren}&=&-\left(\frac{1}{2}\mu^2 + 3\frac{\lambda}{64\pi^2}\mu^2
   + \frac{1}{2}\delta\mu^2\right)v^2\nonumber\\
   &+&\left(\frac{1}{4}\lambda + 3\frac{\lambda^2}{128\pi^2}
   - 6\frac{g^4}{32\pi^2} + \frac{1}{4}\delta\lambda\right)v^4\nonumber\\
   &-& 24\frac{m_q^4}{64\pi^2}\ln\left(\frac{m_q^2}{4}\right) + 
   3\frac{m_\pi^4}{64\pi^2}\ln\left(\frac{m_\pi^2}{4}\right),
\label{Veffrenorsininf}
\ee 
where, hereafter, when not explicitly indicated, the pion and quark masses are
the $v$-dependent ones, given in Eqs.~(\ref{masses}). 

To fix one of the remaining counter-terms, either $\delta\mu^2$ or
$\delta\lambda$, we impose the condition that the minimum of the renormalized
effective potential remains at its classical value, namely,
\be
   \left.\frac{1}{2v}\frac{\partial}{\partial v} V^{(1)}_{ren}\right|_{v=v_0}
   =0,
\label{classmin}
\ee
which implies that 
\be
   \delta\lambda = 6\frac{g^4}{4\pi^2}
   \left[1 + \ln\left(\frac{g^2\mu^2}{4\lambda}\right)\right] + 
   \frac{\lambda}{\mu^2}\delta\mu^2.
\label{deltalambda}
\ee
Inserting Eq.~(\ref{deltalambda}) into Eq.~(\ref{Veffrenorsininf}), we get
\be
   V^{(1)}_{ren}&=&-\left(\frac{1}{2}\mu^2 + 3\frac{\lambda}{64\pi^2}\mu^2
   + \frac{1}{2}\delta\mu^2\right)v^2\nonumber\\
   &+&\left(\frac{1}{4}\lambda + 3\frac{\lambda^2}{128\pi^2}
   + 6\frac{g^4}{32\pi^2} + \frac{\lambda}{4\mu^2}\delta\mu^2\right)v^4
   \nonumber\\
   &-& 6\frac{m_q^4}{16\pi^2}\ln\left(\frac{m_q^2}{m_q^2(v_0)}\right)
   + 3\frac{m_\pi^4}{64\pi^2}\ln\left(\frac{m_\pi^2}{4}\right).\nonumber\\
\label{Vrencasi}
\ee
To fix the second counter-term, $\delta\mu^2$, notice that the argument of the
last logarithmic function is dimensionfull, though the whole term is well
defined as $m_\pi$ goes to zero. However, if we choose
\be
   \delta\mu^2 = -3\frac{\lambda\mu^2}{16\pi^2}
   \ln\left(\frac{\mu^2}{4}\right),
\label{choice}
\ee
we obtain
\be
   V^{(1)}_{ren}&=&-\left(\frac{1}{2}\mu^2 + 3\frac{\lambda}{64\pi^2}\mu^2
   \right)v^2\nonumber\\
   &+&\left(\frac{1}{4}\lambda + 3\frac{\lambda^2}{128\pi^2}
   + 6\frac{g^4}{32\pi^2}\right)v^4
   \nonumber\\
   &-& 6\frac{m_q^4}{16\pi^2}\ln\left(\frac{m_q^2}{m_q^2(v_0)}\right)
   + 3\frac{m_\pi^4}{64\pi^2}\ln\left(\frac{m_\pi^2}{\mu^2}\right),\nonumber\\
\label{Vrenfin}
\ee
which no longer contains dimension-full arguments of the logarithmic
functions. The choice of Eq.~(\ref{choice}) although seemingly arbitrary, has
the advantage of producing a well defined renormalized effective potential
which preserves the properties of the tree level one, namely, that the pions
are massless at $v_0$ which in turn keeps being the minimum of the
potential. Recall that the effective potential is not in itself an observable;
only physical properties extracted from it, such as the position of the
minimum and the critical temperature are. The choice that produces
Eq.~(\ref{Vrenfin}) is also extensively used in Standard Model
calculations~\cite{Carrington}.

We now proceed to include the finite temperature contribution at one-loop. The
finite temperature pieces of Eqs.~(\ref{Vbos1loop}) and~(\ref{Vfer1loop}) 
are given by
\be
   V^{(1)T\neq 0}_{f}&=&6\left[ -\frac{7\pi^2}{180}T^4 +
   \frac{m_q^2}{12}T^2 +
   \frac{m_q^4}{16\pi^2}\ln\left(\frac{m_q^2}{T^2}\right)\right]\nonumber\\
   V^{(1)T\neq 0}_{\pi}&=&3\left[ -\frac{\pi^2}{90}T^4 +
   \frac{m_\pi^2}{24}T^2 - \frac{m_\pi^3}{12\pi}T\right.\nonumber\\
   &-&\left.
   \frac{m_\pi^4}{64\pi^2}\ln\left(\frac{m_\pi^2}{(4\pi T)^2}\right)\right].
   \label{finT}
\ee
Adding the renormalized effective potential to the finite temperature
contributions, we get the full one-loop finite temperature effective potential
\be
   V^{(1)T\neq 0}_{ren}&=&
   -48\frac{\pi^2T^4}{180} - \left( 1 + 3\frac{\lambda}{32\pi^2}\right)
   \frac{\mu^2}{2}v^2\nonumber\\
   &+& \left(\lambda + 3\frac{\lambda^2}{32\pi^2} + 6\frac{g^4}{8\pi^2}\right)
   \frac{v^4}{4}\nonumber\\
   &+&\left(12m_q^2 + 3m_\pi^2\right)\frac{T^2}{24} - 3m_\pi^3\frac{T}{12\pi}
   \nonumber\\
   &-&6\frac{m_q^4}{16\pi^2}\ln\left(\frac{T^2}{m_q^2(v_0)}\right)
   +3\frac{m_\pi^4}{64\pi^2}\ln\left(\frac{(4\pi T)^2}{\mu^2}\right).
   \nonumber\\
\label{VTfinfinal}
\ee
A few words about the properties of Eq.~(\ref{VTfinfinal}) are in order:
First, notice that the dependence on the pion mass in the argument of the
logarithmic functions has canceled upon addition of the vacuum and
finite-temperature pieces of the renormalized effective potential. This is an
important property for otherwise, this function can develop an imaginary part
when the pion mass is negative, namely for $v<v_0$. Secondly, notice the
appearance of a cubic pion mass term. This is also a dangerous term since it
gives rise to an imaginary piece when the pion mass is negative. However, as
we will show, this term is exactly canceled when considering the contribution 
from the ring diagrams.

In order to closely examine the behavior of the renormalized
finite-temperature effective potential with the temperature, let us re-express
Eq.~(\ref{VTfinfinal}) expanding it in powers of $v$
\be
  V^{(1)T\neq 0}_{ren}&=&
   -48\frac{\pi^2T^4}{180} - 3\frac{\mu^2}{24}T^2
   + 3\frac{\mu^4}{64\pi^2}\ln\left(\frac{(4\pi T)^2}{\mu^2}\right)\nonumber\\
   &-&\left[ 1 + 3\frac{\lambda}{32\pi^2}-\frac{2g^2 + \lambda/2}{2\mu^2}T^2
   \right.\nonumber\\
   &+& \left.3\frac{\lambda}{16\pi^2}\ln\left(\frac{(4\pi T)^2}{\mu^2}\right)
   \right]\frac{\mu^2v^2}{2}\nonumber\\
   &+& \left[\lambda + 3\frac{\lambda^2}{32\pi^2} + 6\frac{g^4}{8\pi^2}
   -6\frac{g^4}{4\pi^2}\ln\left(\frac{T^2}{m_q^2(v_0)}\right)\right.
   \nonumber\\
   &+&\left.3\frac{\lambda^2}{16\pi^2}\ln\left(\frac{(4\pi T)^2}{\mu^2}\right)
   \right]\frac{v^4}{4} - 3\frac{m_\pi^3}{12\pi}T.
\label{reexpress}
\ee 

Let us ignore, for the time being, the term proportional to $m_\pi^3$. Notice
that the critical temperature $T_c$ for the phase transition is 
determined by the curvature of the effective potential at $v=0$. When the
curvature changes sign the phase transition starts. At high temperature
the effective potential at $v=0$ is convex  and the minimum of the
potential happens for $v=0$, that is the symmetric phase. However, when the
effective potential becomes concave at $v=0$, the minimum of the effective
potential is located at a finite value of $v$, that is the broken symmetry
phase. The above takes place provided the coefficient of the term 
$v^4$ is positive for otherwise the effective potential for large $v$ becomes
concave and thus unstable. This can happen for very high temperatures larger
than a given temperature $T_{max}$. Therefore, the condition for the analysis
to be valid is that the critical temperature is smaller than this last
temperature, namely, $T_c<T_{max}$.

\begin{figure}[t!] 
{\centering
{\epsfig{file=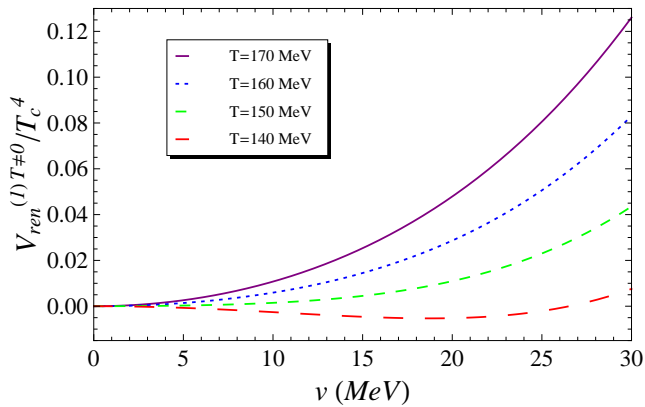, width=1\columnwidth}}
\par}
\caption{$V_{ren}^{(1)T\neq 0}$ scaled by $T_c^4$ and ignoring 
$v$-independent terms as well as the cubic term in the pion mass, for several 
values of the temperature.}
\label{fig1}
\end{figure}

To test whether the analysis is consistent, we proceed to compute these
temperatures. For this purpose, we use the standard values for the parameters
\be
   \lambda &=&20\nn
   \mu &=&380 {\mbox{ MeV}}\nn
   g&=&3.3,
\label{standard}
\ee
which are determined from requiring that the vacuum $\sigma$ mass is,
$m_\sigma^{vac}\simeq 600$ MeV, the constituent quark mass $m_q^{vac}\simeq
300$ MeV and by using a value for the pion vacuum decay constant $f_\pi=93$
MeV. The critical temperature is computed from solving for the temperature 
where the second derivative of the effective potential with respect to $v$
vanishes, the maximum temperature is computed by finding the temperature for
which the coefficient of the quartic term vanishes. This gives
\be
   T_c&=&147.5 {\mbox{ MeV}}\nn
   T_{max}&=&7,196.5 {\mbox{ MeV}},
\label{temps}
\ee
which confirms that $T_c<T_{max}$. The phase transition becomes a smooth
crossover. Figure~\ref{fig1} shows the potential $V_{ren}^{(1)T\neq 0}$,
dropping out $v$-independent terms, scaled by $T_c^4$, for different values of
the temperature and ignoring the cubic term in the pion mass. Notice how this
potential flattens out continuously as the temperature lowers down up to the
temperature where the phase transition takes place, after which, the
potential develops a minimum at a finite value of $v$.

\subsection{Ring diagrams}\label{secIIIb}

\begin{figure}[t!] 
{\centering
{\epsfig{file=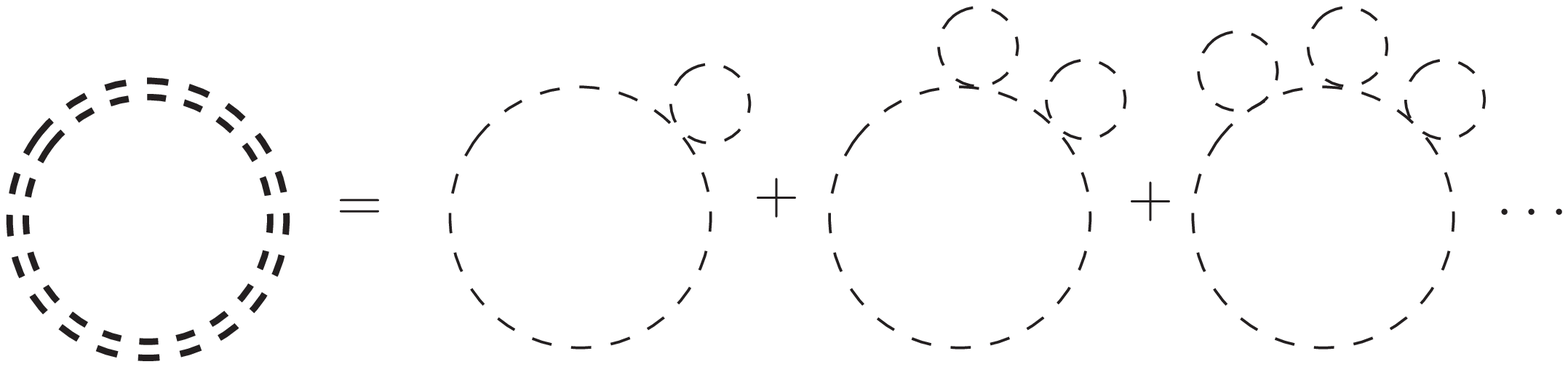, width=0.8\columnwidth}}
\par}
\caption{Schematic representation of the resummation of the ring diagrams.}
\label{fig2}
\end{figure}

We now proceed to include in the analysis the contribution from the ring
diagrams. As is well known, for theories with massless modes, perturbative
calculations can lead to the appearance of infrared divergences which signal
the need of a resummation scheme. The leading divergences can be summed up to
render an infrared safe quantity and the diagrams corresponding to this
divergences are known as the ring diagrams. These are depicted in
Fig.~\ref{fig2} and the explicit expression for the resummed series is given
by
\be
   V^{(ring)}&=&- \frac{T}{2}\sum_n\int\frac{d^3k}{(2\pi )^3}\nn
   &\times&
   \sum_{N=1}^\infty\frac{1}{N}\Big\{2[\Pi^*\Delta^B(k)]^N
   +[\Pi^0\Delta^0(k)]^N\Big\},\nn
   &=&\frac{T}{2}\sum_n\int\frac{d^3k}{(2\pi )^3}\nn
   &\times&
   \Big\{2\ln [1+\Pi^*\Delta^B(k)] + \ln [1+\Pi^0\Delta^0(k)]\Big\},\nn
\label{rings}
\ee
where $\Pi^*$ and $\Pi^0$ are the charged and neutral pion
self-energies in the presence of the magnetic field and $\Delta^B$ and
$\Delta^0$ their corresponding propagators. The factor $2$ takes into account
that there are two charged pions.

At this point we should mention that, as the couplings (expansion parameters)
in the expression for the pion self-energy are too large for a perturbative
calculation to be strictly valid, it is likely that the one loop expansion
overshoots the exact result. For the purposes of this work where we explore
whether a weak magnetic field may have some influence on the dynamics of the
phase transition, we will content ourselves with a qualitative description of
the problem. For this, we parameterize our ignorance on the self-energies
computing them perturbatively up to one-loop order but multiplying these times
a constant $0< c <1$ that we will vary to explore the parameter space. The
correct line of action is to calculate the pion self energy non 
perturbatively. The continuum approach in this connection would be to solve
its Schwinger-Dyson equation consistently in the presence of a heat bath and
uniform external magnetic field, and then substitute the result into
Eqs.~(\ref{rings}). However, in the present context, we replace this
procedure by introducing the free parameter $c$ as mentioned before.

Equation~(\ref{rings}) contains both vacuum as well as $T$-dependent
infinities. These last can be canceled by a one-loop vacuum counterterm that
renormalizes the pion mass. The procedure is best carried out by explicitly
separating the two-loop contribution to Eq.~(\ref{rings}) which results in the
expression~\cite{LeBellac}
\be
   V^{(ring)}&=&\frac{T}{2}\sum_n\int\frac{d^3k}{(2\pi )^3}\nn
   &\times&
   \Big\{2(\ln [1+\Pi^*\Delta^B(k)] - \Pi^*\Delta^B(k) )\nn
   &+& (\ln [1+\Pi^0\Delta^0(k)] - \Pi^0\Delta^0(k) )\Big\}\nn
   &+& 2V^{*(2)} + V^{0(2)}, 
\label{ringseparated}
\ee   
where $V^{*(2)}$ and $V^{0(2)}$ are the contributions to the two-loop
effective potential for a charged and a neutral pion. Once again, the factor
$2$ takes into account that there are two charged pions.

For the charged pion contribution to the one-loop pion self-energy, we use the
charged scalar propagator which in the weak field limit is given
by~\cite{Ayala}
\be
   \Delta^B=\frac{1}{\omega_n^2+E_k^2}\left\{
   1 - \frac{(eB)^2}{(\omega_n^2+E_k^2)^2} +
   \frac{2(eB)^2k_\perp^2}{(\omega_n^2+E_k^2)^3}\right\},
\label{porpchargscalar}
\ee
where $\omega_n=2n\pi T$, with $n$ an integer, $E_k^2=k_3^2+k_\perp^2
+m_\pi^2$, with $k_\perp^2=k_1^2+k_2^2$ and $m_\pi$ the $v$-dependent
pion mass. For the neutral pion contribution to one the loop pion self-energy,
we use the same propagator in Eq.~(\ref{porpchargscalar}) setting $eB=0$. The
quark contribution to the pion self-energies should in principle consider that
quarks are subject to interact with the external magnetic field. Nevertheless,
for the hierarchy of energy scales we work with, it is easy to see that the
$B$-dependent part of this self-energy is subdominant and thus we just consider
the $B$-independent quark propagator.

The dominant contribution in Eq.~(\ref{ringseparated}) comes from the mode
with $n=0$. Fermions do not contribute to the ring diagrams as their mode with
$n=0$ does not vanish. It is easy to check that, after mass renormalization,
the computation of Eq.~(\ref{ringseparated}) (for the mode with $n=0$) reduces
to considering only the $T$-dependent terms.  As it is outlined in
the appendix, carrying out an expansion of the
argument of the logarithms and keeping only terms up to
$\mathcal{O}\ (eB)^2$, the dominant contribution from the ring diagrams,
according to the hierarchy of energy scales we are considering, can 
be written as
\begin{figure}[t!] 
{\centering
{\epsfig{file=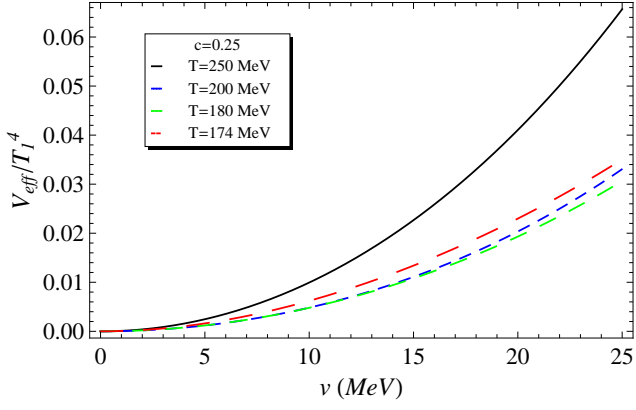, width=1.0\columnwidth}}
\par}
\caption{Effective potential scaled by $T_1(0.25)^4$, ignoring $v$-independent
terms, computed for $\lambda =20$, $\mu =380$ MeV and $g=3.3$ in the absence
of a magnetic field, for several temperatures.}
\label{fig3}
\end{figure}
\be
   V^{(ring)}&=&\frac{5\lambda T^4}{192} +
   \frac{15\lambda}{64\pi^2} T^2m_\pi^2
   + 3\frac{m_\pi^3}{12\pi}T\nn  
   &-&
   \frac{(eB)^2}{192\pi}\frac{\Pi_1T}{(m_\pi^2 + \Pi_1 )^{3/2}}\nn
   &-& 
   \frac{T}{12\pi}(m_\pi^2 + \Pi^0)^{3/2} -
   2\frac{T}{12\pi}(m_\pi^2 + \Pi^*)^{3/2},\nn
\label{Vringfin}
\ee
 where in the fourth term on the right-hand side of
  Eq.~(\ref{Vringfin}) we have kept only the leading contribution 
in $(eB)$. To cope with the lack of a non-perturbative calculation for the
self-energies, from now on we set $\Pi^0$, $\Pi^*$ and $\Pi_1$ as given by
\be
   \Pi^0&=&c\ \Pi^0_{pert}\nn
        &=&c \left[5\lambda\left(\frac{T^2}{12} -
   \frac{(eB)^2}{240\pi}\frac{T}{\tilde{m}_\pi^3}\right) 
   + g^2T^2\right]\nn
   \Pi^*&=&c\ \Pi^*_{pert}\nn
        &=&c \left[5\lambda\left(\frac{T^2}{12} -
   \frac{(eB)^2}{120\pi}\frac{T}{\tilde{m}_\pi^3}\right) 
   + g^2T^2\right]\nn
   \Pi_1&=&c\left[\frac{5\lambda}{12} + g^2\right]T^2,
\label{pis}
\ee
with $0<c<1$ and where, as sketched in the appendix, we make use
of the same cancellations that effectively substitute
$m_\pi\rightarrow\tilde{m}_\pi$, with
\be
   \tilde{m}_\pi=\sqrt{m_\pi^2 + \Pi_1},
   \label{tildem}
\ee
as in the calculation of the ring diagrams. There are several properties of
Eqs.~(\ref{Vringfin}) and~(\ref{pis}) worth 
mentioning: First, notice that, as promised, the cubic  pion mass
term in Eq.~(\ref{Vringfin}) exactly cancels the one appearing in
Eq.~(\ref{reexpress}). Second,  as outlined in the appendix,  the
leading contribution in $(eB)$, in the first and second of Eqs.~(\ref{pis})
need only to consider $\Pi_1$ as a correction to the pion mass.  
%
\begin{figure}[t!] 
{\centering
{\epsfig{file=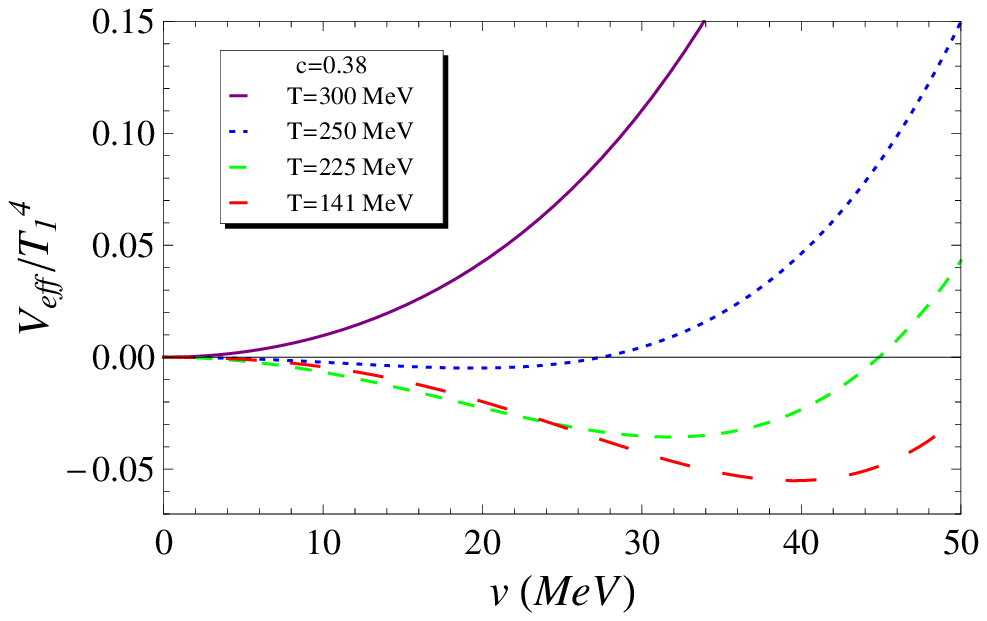, width=1.0\columnwidth}}
\par}
\caption{Effective potential scaled by $T_1(0.38)^4$, ignoring $v$-independent
terms, computed for $\lambda =20$, $\mu =380$ MeV and $g=3.3$ in the absence
of a magnetic field, for serveral temperatures.}
\label{fig4}
\end{figure}

From Eqs.~(\ref{reexpress}) and~(\ref{Vringfin}), the final expression for the
effective potential up to order ring can be written as
\be
   V_{eff}&=&- \left( 1 + 3\frac{\lambda}{32\pi^2}\right)
   \frac{\mu^2}{2}v^2\nn
   &+& \left(\lambda + 3\frac{\lambda^2}{32\pi^2} + 6\frac{g^4}{8\pi^2}\right)
   \frac{v^4}{4}\nn
   &+&\left(12m_q^2 + 3m_\pi^2  +
   \frac{45\lambda}{8\pi^2}m_\pi^2\right) 
   \frac{T^2}{24}\nn
   &-&6\frac{m_q^4}{16\pi^2}\ln\left(\frac{T^2}{m_q^2(v_0)}\right)
   +3\frac{m_\pi^4}{64\pi^2}\ln\left(\frac{(4\pi T)^2}{\mu^2}\right)\nn
   &-& 
   \frac{T}{12\pi}(m_\pi^2 + \Pi^0)^{3/2} -
   2\frac{T}{12\pi}(m_\pi^2 + \Pi^*)^{3/2}\nn  
   &-&
   \frac{(eB)^2}{192\pi}\frac{\Pi_1T}{(m_\pi^2 + \Pi_1 )^{3/2}}
   ,
\label{Vefftot}
\ee
where we have dropped out the explicit $v$-independent terms.
The potential is real provided that $m_\pi^2+\Pi_1>0$ and 
$m_\pi^2+\Pi^*>0$ (this last condition is enough to make sure that
$m_\pi^2+\Pi^0>0$). The first condition holds if it happens
for $v=0$ which yields the requirement that
\be
   T>\frac{\mu}{\sqrt{c(5\lambda T^2/12 + g^2)}}\equiv T_1(c).
\label{T1}
\ee 
The second condition defines a a magnetic field dependent value of a
temperature $T_B$ below which the analysis breaks down. 
\begin{figure}[t!] 
{\centering
{\epsfig{file=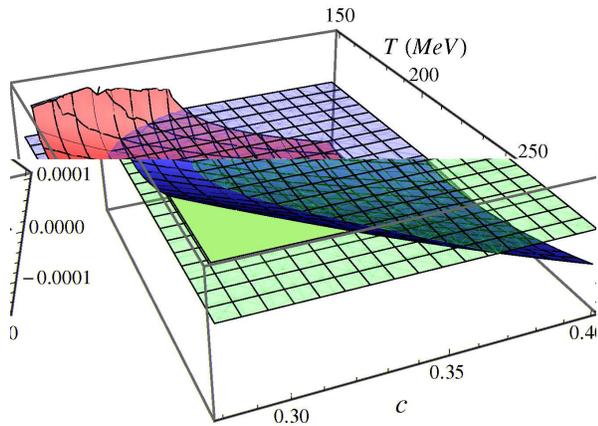, width=0.9\columnwidth}}
\par}
\caption{The second derivative of $V_{eff}$ at $v=0$ as a function of $c$ and
$T$ for $eB=0$.}
\label{fig5}
\end{figure}

\section{Results}\label{secIV}

To explore the properties of the effective potential in Eq.~(\ref{Vefftot}),
we first study the case with $B=0$ and $c=0.25$. Figure~\ref{fig3} shows the
effective potential scaled by $T_1(0.25)^4$, dropping out $v$-independent
terms, computed for the same values of the parameters as in 
Eq.~(\ref{standard}). Notice that for the highest temperature the minimum of
the effective potential is at $v=0$. As the temperature decreases, the
potential flattens at $v=0$ with its curvature remaining positive. However an
interesting phenomenon happens, namely, that for even lower temperatures, the
potential becomes steeper at $v=0$. This behavior persists down to the lowest
temperature allowed. This effect is caused by the ring diagrams {\it even in
the absence of the magnetic field} and thus precludes the phase transition
to occur.  

\begin{figure}[t!] 
{\centering
{\epsfig{file=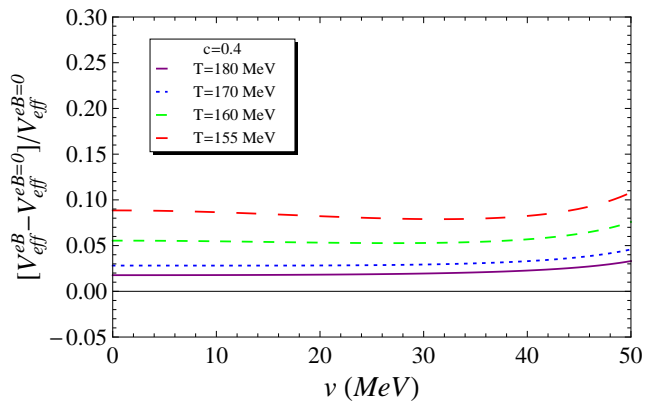, width=1.0\columnwidth}}
\par}
\caption{Relative difference $(V_{eff}^{eB}-V_{eff}^{eB=0})/V_{eff}^{eB=0}$
computed for a field intensity $eB=0.9 (m_\pi^{vac})^2$ and $\lambda =20$,
$\mu=380$ MeV, $g=3.3$.}
\label{fig6}
\end{figure}
To see how this behavior is affected by varying the model parameters, we keep
$B=0$ but change the parameter $c$ which we now take as
$c=0.38$. Figure~\ref{fig4} shows the effective potential scaled by
$T_1(0.38)^4$, dropping out $v$-independent terms, computed for the same values
of the parameters as in Eq.~(\ref{standard}). Notice that for the highest
temperature the minimum of the effective potential is at $v=0$. As the
temperature decreases, the potential flattens and the curvature at $v=0$
becomes negative thus, a minimum at a finite value of $v$
develops. 
This behavior
persists down to the lowest temperature allowed. Thus we see that for certain
values of $c$ the phase transition is allowed but below a certain {\it
critical} value $c_c$, the phase transition does not occur. 
This also indicates that, as the temperature decreases after the
development of the minimum at $v\neq 0$, the system goes back to the symmetry
restored phase.


To find the critical value for the parameter $c$, we can see whether the
condition for the curvature at $v=0$ to change sign is satisfied for real
values of $T$. This is shown in Fig.~\ref{fig5} where we plot the second
derivative of the effective potential at $v=0$ as a function of $c$ and $T$. 
Notice that this second derivative can change sign from positive to negative
as the temperature decreases only above the critical value $c_c\simeq
0.3$. Notice also from Fig.~\ref{fig5} that for a given 
temperature, as the value of $c$ increases the curvature changes sign from
positive to negative. This signals that the system starts out
in the broken symmetry phase requiring a much larger value of the temperature
to restore the symmetry.

We now proceed to analyze the behavior of the effective potential including
the effects of the magnetic field. Figure~\ref{fig6} shows the relative
difference $(V_{eff}^{eB}-V_{eff}^{eB=0})/V_{eff}^{eB=0}$ computed for
the same values of the parameters as in Eq.~(\ref{standard}) and a field
intensity $eB=0.9\times(m_\pi^{vac})^2$. For the chosen temperatures, this
relative difference amounts only for a $5\% - 10\%$ showing that up to this
order in the approximation the effects of the magnetic field are small. The
difference becomes a bit more significant for the lowest temperatures. 

\section{Conclusions}\label{secV}

In conclusion, we have studied the chiral phase transition in the linear sigma
model, including quarks, at finite temperature in the presence of weak magnetic
fields. This has been accomplished by looking at the effective potential up to
the contribution of the ring diagrams, which in the weak field limit, is the
first order where the effects of the magnetic field become evident. 

These diagrams are in principle an 
important ingredient in the calculation, given that the theory contains
massless quantum modes which make it necessary to consider a 
resummation scheme. To accomplish the resummation, one needs to compute the
pion self-energy. Nonetheless, given the fact that the theory has large
coupling constants, the computation of such self-energy is not perturbatively
reliable. In an attempt to qualitatively study the structure of the theory, we
have gone ahead and computed this self-energy perturbatively but have
parameterized our lack of a non-perturbative calculation introducing a factor
$c$ which we allow to vary such that $0 < c < 1$. In these terms, the analysis
shows that considering the ring diagrams, there is a critical value for $c$
below which the phase transition is precluded from happening. Above this
critical value, the phase transition keeps being second order but as the
temperature drops the system comes back to the symmetry restored phase. This
means that if the phase transition was completed, then the bounce back does
not have any effect. However if the transition is delayed, part of the system
could get trapped in the symmetric phase. Finally, the effects of the magnetic
field, in the weak field limit, only account for differences up to about $5\%
- 10\%$ and become larger as the system cools down to the lowest temperatures
for the analysis to be valid. 

In order to be able to extract more reliable conclusions, it is clear that one
needs to have a better control on the calculation of the self-energy for which
a non-perturbative treatment is called for, including the effects of the heat
bath as well as those of the magnetic field. Nevertheless, the analysis shows
in a qualitative way that if this calculation happens to yield values for the
self-energy in the vicinity of the ones parameterized by $c_c$, the magnetic
field could retard the phase transition. This line of thinking is to be pursued
in the future.

\section*{Acknowledgments}

A.A. wishes to thank the kind hospitality of both faculty and staff in
IFM-UMSNH and CBPF during sabbatical visits and the financial support of CNPq,
and DGAPA-UNAM under PAPIIT grant No. IN116008. A.B. and A.R. acknowledge
COECyT, CIC and CONACyT grants. A.S. acknowledges a CONACyT postdoctoral
grant. 

\section*{Appendix: Computation of ring diagrams}

For the sake of the argument, consider only a single charged scalar field. The
generalization to the sigma model is immediate once we consider the isospin
factors. The effective potential at ring-order is given by 
\be
   V^{(ring)}&=&\frac{1}{2}T\sum_n\int\frac{d^3k}{(2\pi)^3}
   \left\{\ln[1+\Pi^B\Delta^B] - \Pi^B\Delta^B\right\}.\nn
   &+& V^{(2)},
\label{A1}
\ee
where $V^{(2)}$ is the two-loop contribution to the effective potential.
Notice that the charged scalar Matsubara propagator is given as in
Eq.~(\ref{porpchargscalar})
and that we can write $\Pi^B=\Pi_1 + \tilde{\Pi}$, where $\Pi_1$ and
$\tilde{\Pi}$ are the magnetic field independent and dependent pieces of
$\Pi^B$. Thus, for small magnetic fields, keeping only the leading terms in
the first equation, we get 
\be
   V^{(ring)}&=&\frac{T}{(2\pi)^2}\int_0^\infty dk k^2
   \left\{\ln\left[1+\frac{\Pi^B}{k^2+m^2}\right.\right.\nn 
   &-&\left.\Pi_1\left(
   \frac{(eB)^2}{(k^2+m^2)^3} -
   \frac{2(eB)^2k_\perp^2}{(k^2+m^2)^4}\right)\right]\nn
   &-&\left[\frac{\Pi^B}{k^2+m^2}\right.\nn
   &-&\left.\left.\Pi_1\left(
   \frac{(eB)^2}{(k^2+m^2)^3} -
   \frac{2(eB)^2k_\perp^2}{k^2+m^2)^4}\right)\right]\right\}\nn
   &+& V^{(2)}.
\label{A2}
\ee
The logarithmic term in the above equation can be written as
\begin{widetext}
\be
   \ln\left[1+\frac{\Pi^B}{k^2+m^2} - \Pi_1\left(
   \frac{(eB)^2}{(k^2+m^2)^3} -
   \frac{2(eB)^2k_\perp^2}{(k^2+m^2)^4}\right)\right]
   &=&
   \ln\left[1+\frac{\Pi^B}{k^2+m^2}\right]\nn
   &+&
   \ln\left[1-
   \frac{(eB)^2\Pi_1}{k^2+m^2+\Pi^B}
   \left(\frac{1}{(k^2+m^2)^2} - \frac{2k_\perp^2}{(k^2+m^2)^3}
   \right)\right]\nn
   &\simeq&
   \ln\left[1+\frac{\Pi^B}{k^2+m^2}\right]\nn
   &-&
   \frac{(eB)^2\Pi_1}{k^2+m^2+\Pi^B}
   \left(\frac{1}{(k^2+m^2)^2} - \frac{2k_\perp^2}{(k^2+m^2)^3}
   \right).
\label{A3}
\ee
\end{widetext}
With this expansion the ring contribution to the effective potential looks
like 
\begin{widetext}
\be
   V^{(ring)}&=&\frac{T}{(2\pi)^2}\int_0^\infty dk k^2
   \left\{\left(\ln\left[1+\frac{\Pi^B}{k^2+m^2}\right]
   -\frac{\Pi^B}{k^2+m^2}\right)-(eB)^2\Pi_1\right.\nn
   &\times&\left.\left[\left(\frac{1}{(k^2+m^2)^2} -
   \frac{2k_\perp^2}{(k^2+m^2)^3}\right)
   \frac{1}{k^2+m^2+\Pi^B}
   - \left(\frac{1}{(k^2+m^2)^2} -
   \frac{2k_\perp^2}{(k^2+m^2)^3}\right)\frac{1}{k^2+m^2}\right]\right\}
   + V^{(2)}.
\label{A4}
\ee
\end{widetext}
In order to render an analytical manageable expression let us make the
approximation in the above equation such that 
\begin{widetext}
\be
   \left(\frac{1}{(k^2+m^2)^2} -
   \frac{2k_\perp^2}{(k^2+m^2)^3}\right)\frac{1}{k^2+m^2+\Pi^B}
   \rightarrow
   \left(\frac{1}{(k^2+m^2+\Pi^B)^2} -
   \frac{2k_\perp^2}{(k^2+m^2+\Pi^B)^3}\right)
   \frac{1}{k^2+m^2+\Pi^B}.
\label{A5}
\ee
\end{widetext}
The error involved is of the order of the leading term in $\Pi$, that is
$\Pi_1$ which could be large if this last is proportional to the coupling
$\lambda$, as is the case in this work. In theories with a small coupling this
error is small. In the absence of a non-perturbative calculation of $\Pi_1$ we
will restrict ourselves to this approximation to keep track of the analytical
structure of the effective potential. 

Thus, under this approximation, we get
\be
   V^{(ring)}&=&\frac{T}{4\pi}\left\{\left[
   \frac{m^3}{3} - \frac{(m^2+\Pi^B)^{3/2}}{3} + \frac{m\Pi^B}{2}\right]
   \right.\nn
   &+&\left.\frac{(eB)^2\Pi_1}{48}\left[\frac{1}{m^3} - 
   \frac{1}{(m^2+\Pi^B)^{3/2}}\right]\right\}\nn
   &+& V^{(2)}.
   \label{A6}
\ee
We now recall that the two-loop contribution to the effective potential
contains a term that can be written as 
\be
   V^{(2)}&\rightarrow&\frac{\lambda}{8}\left[
   \frac{2\Pi_1}{\lambda} - \frac{(eB)^2T}{96\pi m^3}\right]^2\nn
   &\simeq&\frac{\lambda}{8}\left[
   \frac{4\Pi_1^2}{\lambda^2} - \frac{(eB)^2T\Pi_1}{24\pi m^3\lambda}\right]
   \nn
   &=&\frac{\lambda}{2}\left[\frac{T^4}{(24)^2} + \frac{m^2T^2}{64\pi^2}
   - \frac{mT^3}{96\pi}\right] - \frac{(eB)^2T\Pi_1}{192\pi m^3},\nn
\label{A7}
\ee
where in the last line we have used that for this theory, up to one-loop order
we have 
\be
   \Pi_1=\lambda\left(\frac{T^2}{24} - \frac{mT}{8\pi}\right).
\label{A8}
\ee
Finally, recall that the one-loop contribution to the effective potential
contains a term that looks like 
\be
   V^{(1)}\rightarrow -\frac{m^3T}{12\pi}.
\label{A9}
\ee
By adding $V^{(1)}+V^{(ring)}$ we see that the potentially dangerous
terms with odd powers of $m$ all cancel and that this cancellation effectively
amounts for the replacement $m\rightarrow\sqrt{m^2+\Pi^B}$. The same
discussion applies in the case of the self-energy as can easily be checked.

\end{document}